%% file: etab_om.tex
\documentclass[%
 reprint,
 nofootinbib,
 amsmath,amssymb,
 aps]{revtex4-2}

\usepackage{graphicx}
\usepackage{dcolumn}
\usepackage{bm}
\usepackage{hyperref}
\usepackage{cleveref}
\usepackage{xcolor}
\usepackage{tabu}

\usepackage{booktabs}
\usepackage{orcidlink}

\DeclareMathAlphabet{\pazocal}{OMS}{zplm}{m}{n}

\newcommand{\mev}{\mathrm{MeV}}
\newcommand{\mevc}{\mathrm{MeV}/c}
\newcommand{\mevm}{\mathrm{MeV}/c^2}
\newcommand{\gev}{\mathrm{GeV}}

\newcommand{\gevm}{\mathrm{GeV}/c^2}

\newcommand{\fb}{\mathrm{fb}^{-1}}

\newcommand{\ee}{e^+e^-}

\newcommand{\pp}{\pi^+\pi^-}
\newcommand{\pn}{\pi^0}
\newcommand{\om}{\omega}
\newcommand{\U}{\Upsilon}
\newcommand{\Uo}{\Upsilon(1S)}
\newcommand{\chbz}{\chi_{b0}(1P)}
\newcommand{\chbo}{\chi_{b1}(1P)}
\newcommand{\chbt}{\chi_{b2}(1P)}
\newcommand{\et}{\eta_b(1S)}
\newcommand{\mrppp}{M_\mathrm{recoil}(\pi^+\pi^-\pi^0)}

\begin{document}

\preprint{APS/123-QED}

\title{Search for the $e^+e^-\to\eta_{b}(1S)\omega$ and
    $e^+e^-\to\chi_{b0}(1P)\omega$ processes at
    $\sqrt{s}=10.745\,\mathrm{GeV}$}

\include{pub032-orcid}

\date{December 20, 2023}

\begin{abstract}
  \noindent
We search for the $e^+e^-\to\eta_b(1S)\omega$ and
$e^+e^-\to\chi_{b0}(1P)\omega$ processes at a center-of-mass energy of
10.745\,GeV, which is close to the peak of the $\Upsilon(10753)$
state. We use data collected by the Belle~II experiment during a
special run, corresponding to an integrated luminosity of
$9.8\,\mathrm{fb}^{-1}$. We reconstruct $\omega\to\pi^+\pi^-\pi^0$
decays and use the $\omega$ meson's recoil mass to search for the
signals. We do not find evidence for either process, and set upper
limits on the corresponding Born-level cross sections of 2.5\,pb and
7.8\,pb, respectively, at the 90\% confidence level. The
$\chi_{b0}(1P)\omega$ limit is the result of a combination of this
analysis and a previous search using full reconstruction.
\end{abstract}

\maketitle

\section{Introduction}

Recently, the Belle experiment observed a new state, the
$\Upsilon(10753)$, as a narrow enhancement in the
$e^+e^-\to\Upsilon(1S,2S,3S)\pi^+\pi^-$ cross
sections~\cite{Belle:2019cbt}. Subsequently, the Belle~II experiment
observed a similar structure in the $e^+e^-\to\chi_{b{1,2}}(1P)\omega$
cross sections, which confirms the existence of the $\Upsilon(10753)$
in additional decay channels~\cite{Belle-II:2022xdi}. 
This state has been interpreted in several ways: as an $\Upsilon(3D)$
bottomonium level mixed with nearby $S$-wave states via hadron
loops~\cite{Badalian:2009bu,Li:2019qsg,Liang:2019geg,Giron:2020qpb}, a
hadronic molecule with a small admixture of
bottomonium~\cite{Bicudo:2020qhp}, a hybrid
meson~\cite{TarrusCastella:2019lyq,Chen:2019uzm,Brambilla:2019esw},
or a compact tetraquark with a diquark-antidiquark
structure~\cite{Wang:2019veq,Ali:2019okl}.
Further studies of hadronic transitions from the $\Upsilon(10753)$ to
lower bottomonia will help to understand its
structure~\cite{Giron:2020qpb,Brambilla:2019esw,Wang:2019veq,Ali:2019okl,TarrusCastella:2021pld,Li:2021jjt,Bai:2022cfz,Li:2022leg,Liu:2023rhh}.

In this paper, we report on a search for the processes $\ee\to\et\om$
and $\ee\to\chbz\om$ at a center-of-mass (c.m.) energy of
$10.745\,\gev$, which is close to the peak of the $\Upsilon(10753)$
state. We use Belle~II data corresponding to an integrated luminosity
of $9.8\,\fb$~\cite{Belle-II:2019usr} collected during a special run
of the SuperKEKB collider at energies above the $\U(4S)$ resonance.

The $\et$ and $\chbz$ mesons do not have exclusive decay channels with
a large product of efficiency and branching fraction. Thus, we
reconstruct only an $\om$ meson in the $\pp\pn$ decay and use recoil
mass,
\begin{equation}        
\label{eq:mrecoil_definition}
\mrppp = \sqrt{\left(\frac{\sqrt{s}-E_{\omega}}{c^2}\right)^2-\left(\frac{p_{\omega}}{c}\right)^2},
\end{equation}
as the signal-extraction variable, where $E_{\omega}$ and $p_{\omega}$
are the energy and momentum of the $\pp\pn$ combination in the
c.m.\ frame.

In a previous study~\cite{Belle-II:2022xdi}, we searched for the
process $\ee\to\chbz\om$ fully reconstructing the $\chbz\to\Uo\gamma$
decay and found no significant signal. The probability of the decay
$\chbz\to\Uo\gamma$ is small, thus, the sensitivity of partial
reconstruction, applied in this analysis, might be higher than that of
full reconstruction.

We follow a blind approach, i.e., the analysis procedure is
established before examining the signal-variable distribution in data.

\section{Belle II detector and simulation}

The analysis is based on a data sample collected with the Belle~II
detector~\cite{Belle-II:2010dht} at the asymmetric-energy $e^+e^-$
SuperKEKB collider~\cite{Akai:2018mbz}.

The detector has a cylindrical geometry with $z$ axis approximately
coincident with the electron beam direction, which defines the forward
direction. Belle II includes a two-layer silicon-pixel detector (PXD)
surrounded by a four-layer double-sided silicon-strip detector and a
56-layer central drift chamber (CDC). These detectors reconstruct
tracks (trajectories of charged particles). Only one sixth of the
second layer of the PXD was installed for the data analyzed here. 
Surrounding the CDC is a time-of-propagation counter (TOP) in the
central region, and an aerogel-based ring-imaging Cherenkov counter
(ARICH) in the forward region. These detectors provide
charged-particle identification. Surrounding the TOP and ARICH is an
electromagnetic calorimeter (ECL) made with CsI(Tl) crystals that
provides energy and timing measurements for photons and
electrons. 
These sub-systems are surrounded by a superconducting solenoid,
providing an axial magnetic field of 1.5~T. An iron flux return
located outside the coil is instrumented with resistive plate chambers
and plastic scintillators to detect $K^0_L$ mesons and to identify
muons (KLM). 

We generate signal events using EvtGen~\cite{Lange:2001uf}. Events are
generated with a uniform distribution over angular variables and then
are weighted according to the theoretical expectations described in
Appendix~\ref{appendix:ang_distr}.
Initial-state radiation (ISR) is simulated using the PHOKHARA
generator~\cite{Rodrigo:2001kf} where processes at next-to-leading
order in the electromagnetic coupling are taken into account. We
assume that the signal processes proceed via the $\U(10753)$ state
with parameters from Ref.~\cite{Belle:2019cbt} and weight events
accordingly.
The GEANT4 package is used to simulate the passage of the particles
inside the detector and its response~\cite{GEANT4:2002zbu}. All the
data and simulated events are reconstructed and analyzed using the
Belle II analysis software~\cite{Kuhr:2018lps,basf2-zenodo}.

\section{Event selection}
\label{sec:selection}

We reconstruct the $\omega$ meson using the $\omega\to\pp\pn$ decay.
Online event selection is based on the number of charged particles and
observed energy in an event, and is fully efficient for signal.
For charged pions, we require the distance from the beam-spot to be
within 0.5\,cm along the $z$ axis and 0.3\,cm in the transverse plane.
Particle identification uses the $dE/dx$ measurements in the CDC and
information from TOP, ARICH, ECL, and KLM. The corresponding
likelihoods are calculated for each particle hypothesis, and loose
selections are applied to separate charged pions from kaons, protons,
and electrons.

Candidate $\pi^0$ mesons are reconstructed from pairs of photons,
which are energy deposits (clusters) in the ECL not matched to a track
in the CDC.
The photon energy is required to exceed $50\,\mev$ in the forward
endcap ($12.4^\circ<\theta<31.4^\circ$) and barrel
($32.2^\circ<\theta<128.7^\circ$) regions of the ECL, and $75\,\mev$
in the backward endcap ($130.7^\circ<\theta<155.1^\circ$), because the
latter region has higher beam-induced background.
To suppress beam-induced background, we require the difference between
the cluster time and the collision time to satisfy
$|\Delta{t}|<50\,\mathrm{ns}$. This requirement corresponds to
approximately two standard deviations in time resolution.
To suppress background from hadronic clusters, which are broader than
electromagnetic ones, we require the ratio of the energy deposit in a
$3\times3$ matrix of crystals to that in the enclosing $5\times5$
matrix in which the four corner crystals are excluded to be greater
than 0.8.
The invariant mass of the selected photon pairs $M(\gamma\gamma)$ is
required to satisfy $|M(\gamma\gamma)-m_{\pi^0}|<12\,\mevm$ for the
$\et\om$ channel and $|M(\gamma\gamma)-m_{\pi^0}|<13\,\mevm$ for the
$\chbz\om$ channel, where $m_{\pi^0}$ is the $\pi^0$
mass~\cite{Workman:2022ynf}. These requirements correspond to
approximately two standard deviations in $\pi^0$ mass resolution. To
suppress combinatorial background from low-energy photons, we require
the momentum of the $\pi^0$ candidate in the c.m.\ frame to exceed
$260\,\mevc$ for the $\et\om$ channel and $130\,\mevc$ for the
$\chbz\om$ channel. We perform a mass-constrained kinematic fit for
the $\pi^0$ candidates to improve the $\pi^0$ momentum resolution.

The $\omega$ candidates are selected by combining $\pi^+$, $\pi^-$,
and $\pi^0$, and requiring the invariant mass to satisfy
$|M(\pi^+\pi^-\pi^0)-m_\omega|<13\,\mevm$, where $m_\omega$ is the
$\omega$ mass~\cite{Workman:2022ynf}.
In the Dalitz plot (DP) of the $\omega\to\pi^+\pi^-\pi^0$ decay, the
density of the signal decreases towards the boundaries, while the
combinatorial background is concentrated near the boundaries. To
suppress the combinatorial background, we use a normalized distance
$r$ to the center of the symmetrized DP~\cite{Belle:2015fvz} in our
candidate selection criteria. The variable $r$ takes the value $r=0$
at the DP center and $r=1$ at its boundary. We require $r<0.84$ for
the $\et$ and $r<0.82$ for the $\chbz$.

To suppress background from continuum events $e^+e^- \to q\bar{q}$
($q=u,d,s,c$), which have a jet-like shape, we use the ratio $R_2$ of
the second to the zeroth order Fox Wolfram
moments~\cite{Fox:1978vu}. For $b\bar{b}$ events, this variable peaks
at approximately 0.1, while for continuum events it is distributed
almost uniformly between 0.0 and 1.0. The selection criteria are
$R_2<0.21$ for the $\et$ and $R_2<0.28$ for the $\chbz$.

The above selection criteria are obtained by maximizing the figure of
merit (FoM), defined as $\epsilon_\mathrm{sig}/\sqrt{N_\mathrm{bkg}}$,
where $\epsilon_\mathrm{sig}$ is the signal efficiency determined
using simulation, and $N_\mathrm{bkg}$ is the background yield
estimated using data. This definition of the FoM takes into account
the small signal-to-background ratio for partial reconstruction. To
find the global maximum of the FoM, we scan each selection variable in
an iterative way.

The $\mrppp$ fit intervals are $(9200,9600)\,\mevm$ and
$(9780,9950)\,\mevm$ for the $\et$ and $\chbz$ candidates,
respectively. For reference, the $\et$ mass is
$(9398.7\pm2.0)\,\mevm$, and the $\chi_{bJ}(1P)$ $(J=0,1,2)$ masses
are $(9859.4\pm0.5)\,\mevm$, $(9892.8\pm0.4)\,\mevm$, and
$(9912.2\pm0.4)\,\mevm$, respectively~\cite{Workman:2022ynf}. The
$\mrppp$ resolutions are $15.0\,\mevm$ for the $\et$ and $8.4\,\mevm$
for the $\chbz$.

There are on average 1.6 and 4.0 candidates per event for the
$\ee\to\et\omega$ and $\ee\to\chbz\om$ processes, respectively. We
verify that the additional candidates do not peak in the $\mrppp$
signal regions, and therefore, we retain all candidates. The
reconstruction efficiencies are 7.6\% and 7.8\% for the $\et\om$ and
$\chbz\om$ channels, respectively. These include corrections for the
discrepancies between data and simulation, which are discussed along
with the systematic uncertainties in Section~\ref{sec:xsec}.

\begin{figure}[htbp]
\begin{center}
\includegraphics[height=11cm, keepaspectratio]{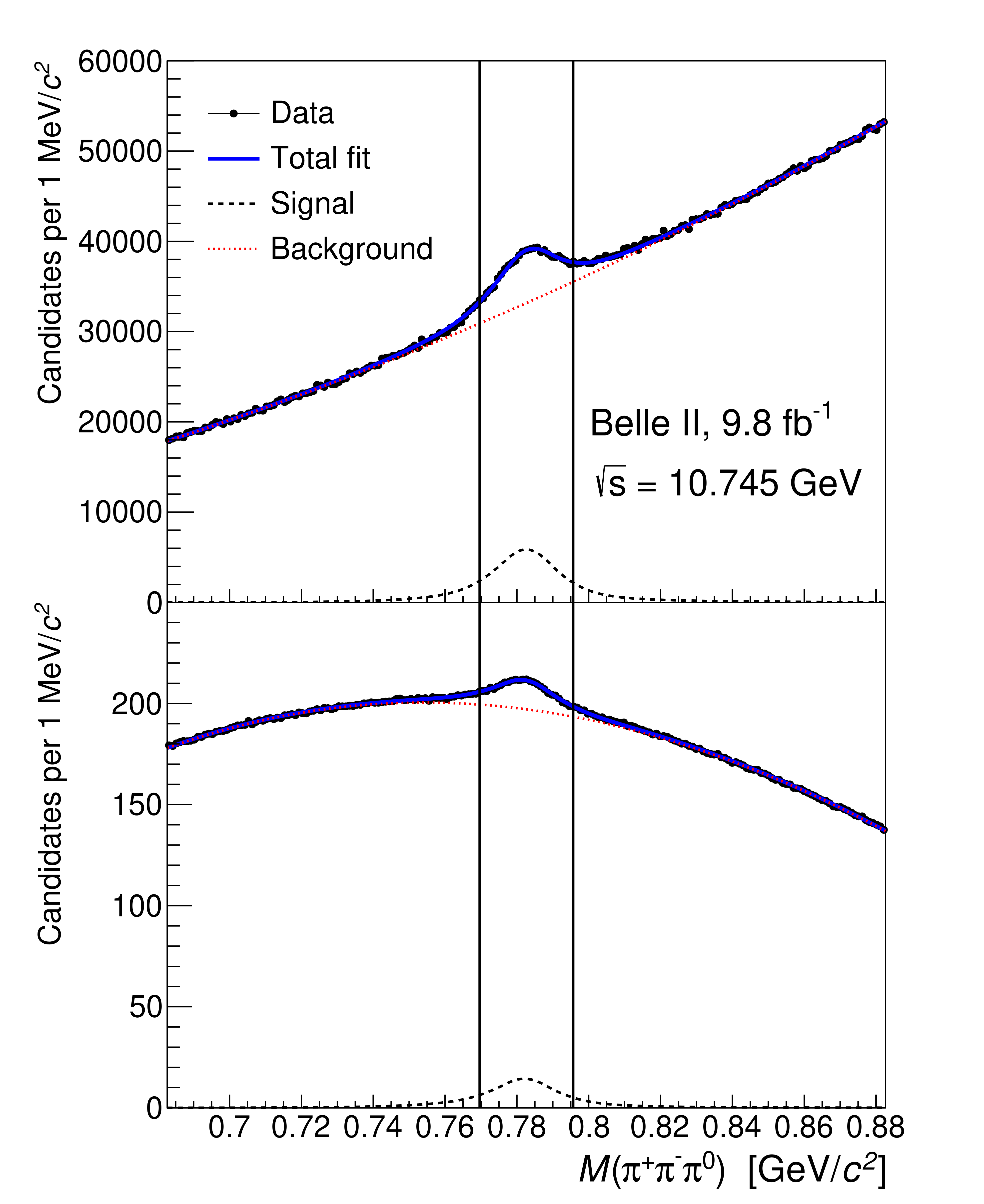}
\caption{\label{fig:etab_omega_fit} Distributions of
  $M(\pi^+\pi^-\pi^0)$ for the $e^+e^-\to\eta_b(1S)\omega$ (top) and
  $e^+e^-\to\chi_{b0}(1P)\omega$ (bottom) candidates. Points indicate
  the data; solid curves show the results of the fit; dashed and
  dotted curves show the signal and background components of the fit,
  respectively. Vertical lines indicate the $\om$ signal region. }
\end{center}
\end{figure}

The $M(\pi^+\pi^-\pi^0)$ distributions for the $\ee\to\et\om$ and
$\ee\to\chbz\om$ candidates without the $\pi^+\pi^-\pi^0$ invariant
mass requirement are shown in Fig.~\ref{fig:etab_omega_fit}.
We perform a least-squares fit to these distributions, in which the
$\omega$-signal shape is modeled by a sum of a Gaussian function and a
double-sided Crystal Ball (CB) function~\cite{Oreglia:1981fx}, while
the background is described by second and third order Chebyshev
polynomials for the $\et\om$ and $\chbz\om$ channels, respectively.
The parameters of the signal fit function are determined from
simulation. To account for possible data-simulation discrepancies, we
introduce an overall mass shift and a width scale factor that are
determined from the fit. We find, respectively, $(0.80\pm0.16)\,\mevm$
and $1.07\pm0.03$ for the $\et$, $(0.17\pm0.18)\,\mevm$ and
$0.99\pm0.03$ for the $\chbz$.
The purity of the $\omega$-meson candidates is estimated to be 12.9\%
and 5.3\% for the $\et\om$ and $\chbz\om$ channels, respectively.

\section{Yield measurement}

\begin{figure}[htbp]
\includegraphics[height=8.5cm, keepaspectratio]{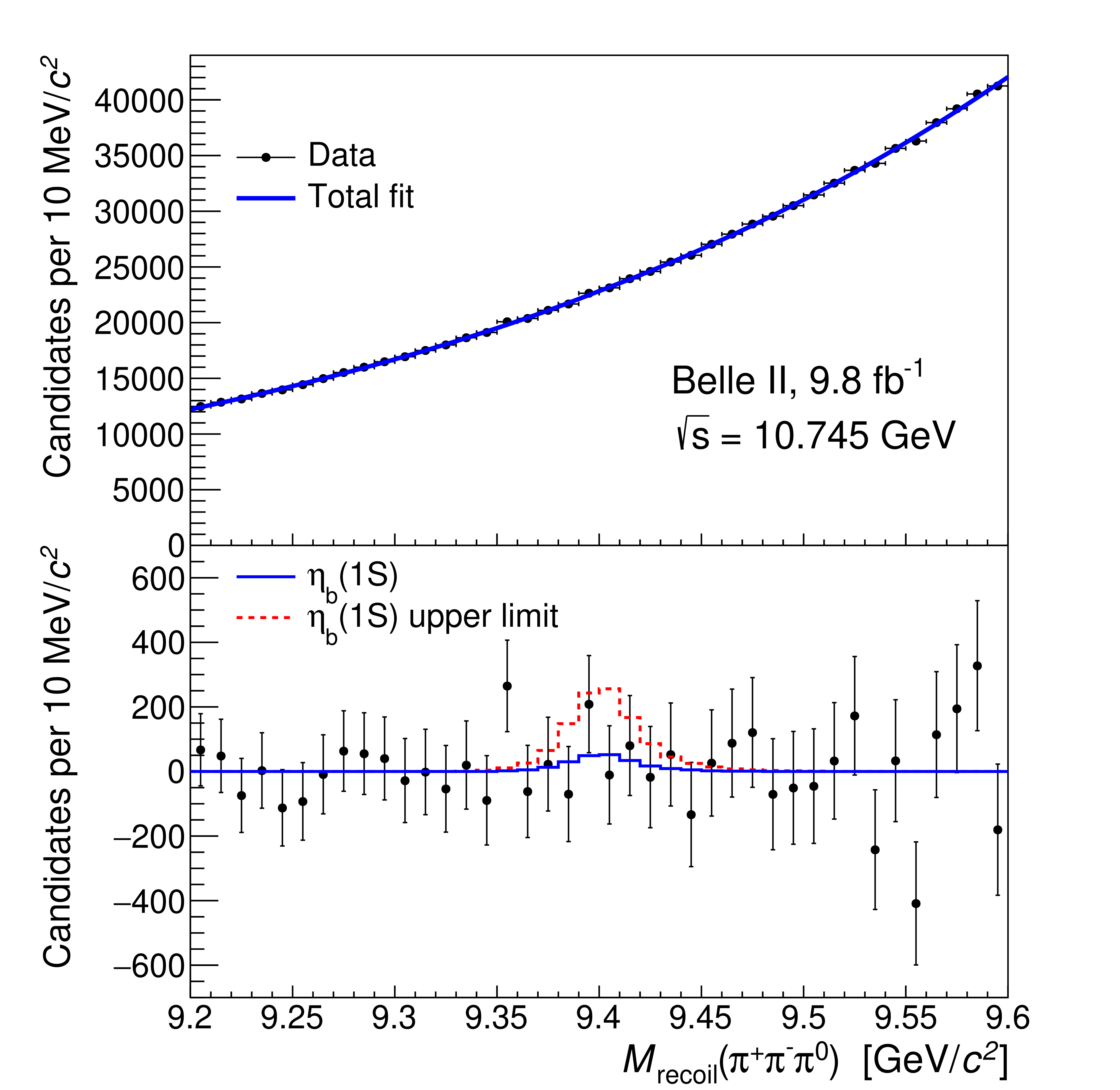}
\caption{\label{fig:etab_data_fit} Distribution of $\mrppp$ for the
  $\ee\to\et\om$ candidates. Top: data points with the fit function
  overlaid. Bottom: the same distributions with the background
  component of the fit function subtracted. The solid histogram shows
  the fit function for the best fit; the dashed histogram shows the
  same function with the yield fixed to the upper limit.}
\end{figure}

\begin{figure}[htbp]
\includegraphics[height=8.5cm, keepaspectratio]{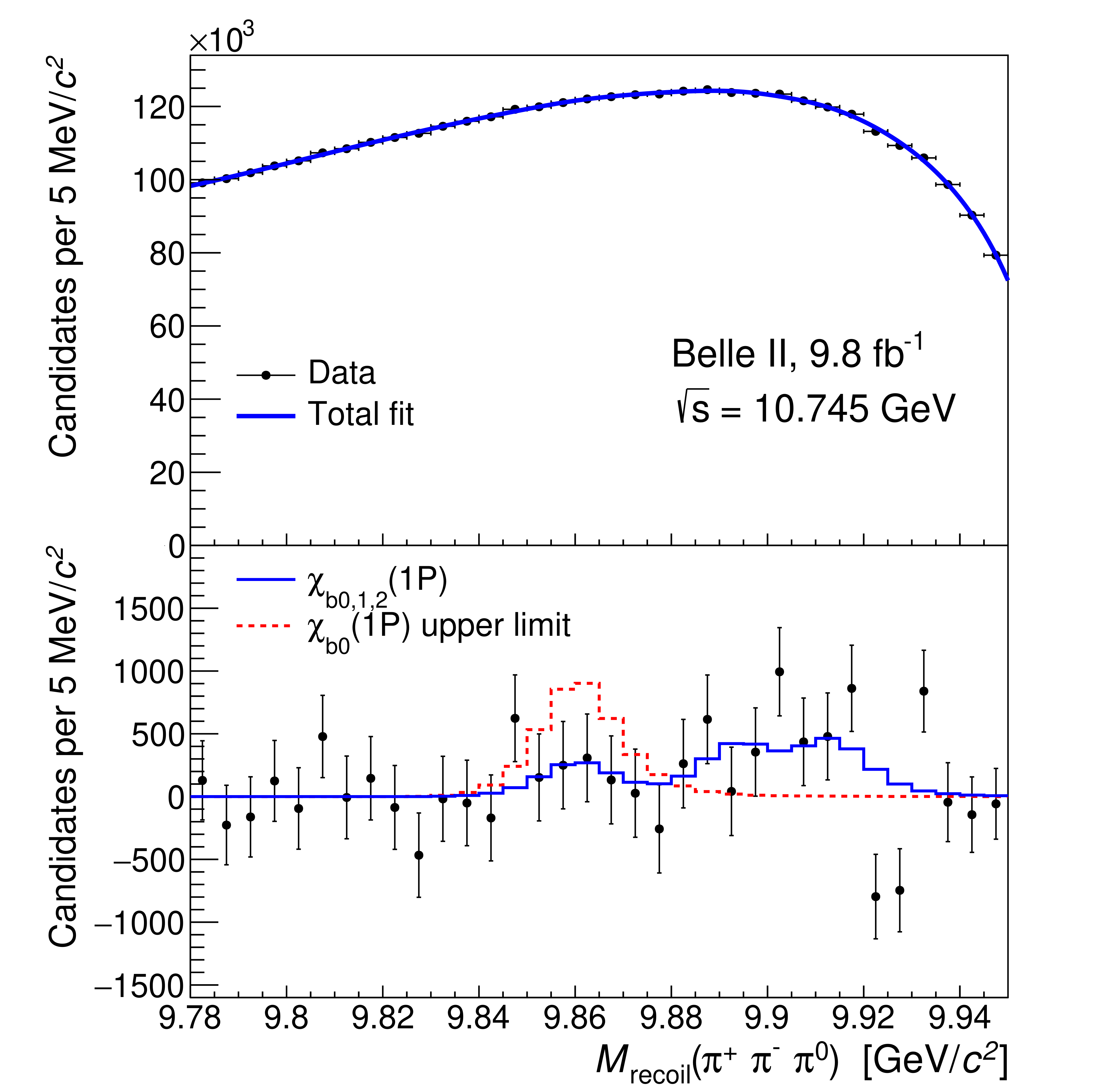}
\caption{\label{fig:chib_data_fit} Distribution of $\mrppp$ for the
  $\ee\to\chbz\om$ candidates.  Symbols are the same as those in
  Fig.~\ref{fig:etab_data_fit}. The $\chbo$ and $\chbt$ contributions
  between 9.88 and $9.94\,\gevm$ are discussed in the text. }
\end{figure}

The $M_\mathrm{recoil}(\pi^{+}\pi^{-}\pi^{0})$ distributions for the
$e^+e^- \to \eta_b(1S)\omega$ and $e^+e^- \to \chi_{b0}(1P)\omega$
candidates are shown in Figs.~\ref{fig:etab_data_fit} and
\ref{fig:chib_data_fit}, respectively. We perform a $\chi^2$ fit
to these distributions, in which the shapes of the $\eta_b(1S)$ and
$\chi_{bJ}(1P)$ signals are fixed to the simulation results. To model
the background, we use a 3rd-order Chebyshev polynomial for the
$\eta_b(1S)$ channel, and the product of a 4th-order Chebyshev
polynomial and a square-root function for the $\chi_{b0}(1P)$
channel. Orders of the polynomial functions are chosen to give the
maximal $p$-value for the fit.

Based on the results of the full-reconstruction
analysis~\cite{Belle-II:2022xdi}, we find that the expected ratio of
the $\chi_{b1}(1P)$ and $\chi_{b2}(1P)$ yields with partial
reconstruction is $N_1/N_2=1.4\pm0.7$. In an initial fit to the data
we fix $N_1/N_2=1.4$, and find $N_1+N_2=(5.5\pm3.2)\times10^3$, which
agrees with the expectation based on Ref.~\cite{Belle-II:2022xdi} of
$(3.4\pm1.0)\times10^3$. In the following, we fix $N_1+N_2$ to the
expected value, which helps to improve the sensitivity to the
$\chi_{b0}(1P)$ signal. Thus, only the $\et$ and $\chbz$ yields, and
background parameters, are free in the fit. The fit results are shown
in Figs.~\ref{fig:etab_data_fit} and \ref{fig:chib_data_fit}. We use
$1\,\mevm$ bins for fitting and $10\,\mevm$ or $5\,\mevm$ bins for
visualization. No significant signals are observed; the obtained $\et$
and $\chbz$ yields are given in Table~\ref{table:yield_XS_UL_data}.

\renewcommand{\arraystretch}{1.1}
\begin{table}[htbp]
\caption{Signal yields and Born-level cross sections (central values and
  upper limits at the 90\% CL) for the processes
  $e^+e^-\to\eta_b(1S)\omega$ and $e^+e^-\to\chi_{b0}(1P)\omega$.}
\begin{tabular}{@{}lcc@{}}
\toprule
 & $\eta_b(1S)\omega$ & $\chi_{b0}(1P)\omega$ \\
\midrule
Yield ($10^3$) & $0.23\pm0.49\pm0.25$ & $1.2\pm1.4\pm0.9$ \\
Born cross section (pb)\;\; & $0.5\pm1.1\pm0.6$ & $2.6\pm3.1\pm2.0$ \\
Upper limit (pb) & $<2.5$ &  $<8.7$ \\
\bottomrule
\end{tabular}
\label{table:yield_XS_UL_data}
\end{table}
\renewcommand{\arraystretch}{1}

\section{Born cross sections and systematic uncertainties}
\label{sec:xsec}

The Born-level cross sections are calculated as
\begin{equation}
    \sigma_\mathrm{B} = \frac{N \, |1 - \Pi|^2}{\epsilon \,
      \mathcal{L} \, (1 + \delta_\mathrm{ISR}) \,
      \mathcal{B}_\mathrm{int}}, 
\end{equation}
where $N$ is the signal yield, $\epsilon$ is the reconstruction
efficiency, $\mathcal{L}$ is the integrated luminosity;
$|1-\Pi|^2=0.93$ is the vacuum
polarization~\cite{WorkingGrouponRadiativeCorrections:2010bjp},
$\mathcal{B}_\mathrm{int}$ is a product of the
$\mathcal{B}(\omega\to\pi^+\pi^-\pi^0)$ and
$\mathcal{B}(\pi^0\to\gamma\gamma)$ branching fractions,
$(1+\delta_\mathrm{ISR})=0.63$ is the radiative correction calculated
using the Kuraev-Fadin radiator~\cite{Kuraev:1985hb} assuming
production via $\Upsilon(10753)$.

The dominant contributions to the systematic uncertainty in the yields
are listed in Table~\ref{table:add_syst}. We vary the $\eta_b(1S)$ and
$\chi_{b0}(1S)$ masses by one standard deviation around their
values~\cite{Workman:2022ynf}. The effect of variation of the
$\eta_b(1S)$ width is negligibly small. We vary the c.m.\ energy in
Eq.~\eqref{eq:mrecoil_definition} by $\pm1\,\mev$ to account for the
uncertainty in its calibration~\cite{Belle-II:BBscan}. We use the
shifts and scale factors of the $M(\pi^+\pi^-\pi^0)$ fits, determined
in Section~\ref{sec:selection}, to calibrate the momentum resolution
in the signal simulation; the resulting change of the yields is
negligibly small. The peak position and the ISR tail of the signal
function depend on the shape of the cross section as a function of
collision energy. We assume that the signal cross section is constant
in energy instead of considering the resonant production via
$\Upsilon(10753)$. Since $N$ and $(1+\delta_\mathrm{ISR})$ are
correlated when the shape of the cross-section is varied, we calculate
the deviation of the ratio $N/(1+\delta_\mathrm{ISR})$. For the
$\chi_{b0}(1P)\omega$ channel, we vary the expected $\chi_{b1}(1P)$
and $\chi_{b2}(1P)$ yields $N_1=(1.9\pm0.4)\times10^3$ and
$N_2=(1.4\pm0.6)\times10^3$ according to their uncertainties
considering their $-0.57$ correlation~\cite{Belle-II:2022xdi}. To
estimate the contribution from the assumptions on the background
shape, we vary both boundaries of the $\eta_b(1S)$ fit interval
simultaneously by $\pm50\,\mevm$, and the lower boundary of the
$\chi_{b0}(1P)$ fit interval by $\pm50\,\mevm$. We also increase the
polynomial order by one. For the background-shape source, we consider
the root-mean-square spread of the deviations to be the corresponding
systematic uncertainty. For other sources, we use maximal
deviations. The total systematic uncertainty in the yields, shown in
Table~\ref{table:add_syst}, is estimated as a sum of the various
contributions in quadrature.

\renewcommand{\arraystretch}{1.2}
\begin{table}[htbp]
\caption{Systematic uncertainties in the yields for the processes
  $\ee\to\et\om$ and $\ee\to\chbz\om$ (in units of $10^3$).}
\centering
\begin{tabular}{@{}lcc@{}} 
\toprule
& \,$\eta_{b}(1S)\omega$\, & \,$\chi_{b0}(1P)\omega$\, \\ 
\midrule
$\eta_{b}(1S)$ / $\chi_{b0}(1P)$ mass & $0.05$ & {$0.08$} \\
Collision-energy calibration & $0.02$ & $0.19$ \\
Cross-section shape & $0.01$  & $0.13$ \\
$\chi_{b1}(1P)$ and $\chi_{b2}(1P)$ yields & $-$  & {$0.27$} \\
Background shape & $ 0.24$ & $ 0.85$ \\
\midrule
Total & $ 0.25$ & $ 0.92$ \\
\bottomrule
\end{tabular}
\label{table:add_syst}
\end{table}
\renewcommand{\arraystretch}{1.0}

A summary of the multiplicative uncertainties is presented in
Table~\ref{table:mult_syst}. The possible discrepancies between data
and simulation contribute to the uncertainty in the reconstruction
efficiency. The uncertainty in track reconstruction efficiency is
estimated using the $\bar{B}^0\to{D}^{*+}(\to{D}^0\pi^+)\pi^-$ decays
for low-momentum particles and using the $\ee\to\tau^+\tau^-$
process for mid-to-high momentum particles.
The uncertainty in the efficiency of the PID requirements is
determined using ${K^{0}_{\rm S}}\to\pi^+\pi^-$,
$D^{*+}\to{}D^0(\to{}K^-\pi^+)\pi^+$, and $\Lambda^0\to{}p\pi^-$
data samples.
The uncertainty in the $\pi^0$ reconstruction efficiency is estimated
by comparing reconstructed $\eta\to\pi^0\pi^0\pi^0$,
$\eta\to\pi^+\pi^-\pi^0$, and $\eta\to\gamma\gamma$ decays.
The efficiencies of the $R_2$ requirements in the $\et\om$ and
$\chbz\om$ channels are 82\% and 91\%, respectively. We assume that
the relative uncertainty in these efficiencies is 10\%. The
uncertainty in luminosity is measured using Bhabha and $\gamma\gamma$
events~\cite{Belle-II:2019usr}. The uncertainties in the
$\omega\to\pi^+\pi^-\pi^0$ and $\pi^0\to\gamma\gamma$ branching
fractions are taken from Ref.~\cite{Workman:2022ynf}.

\begin{table}[htbp]
\caption{Multiplicative systematic uncertainties for the measurement
  of the $\ee\to\et\om$ and $\ee\to\chbz\om$ cross sections (in \%).} 
\centering
\begin{tabular}{lcc} 
\toprule
& \,$\eta_{b}(1S) \omega$\, & \,$\chi_{b0}(1P) \omega$\, \\ 
\midrule
Track reconstruction efficiency & 1.6 & 2.4  \\
PID efficiency & 0.8 & 1.0  \\
$\pi^{0}$ reconstruction efficiency & 3.2 & 7.3 \\
{$R_2$ efficiency} & 10.0 & 10.0 \\
Luminosity & 0.6 & 0.6 \\
$\mathcal{B}(\om\to\pp\pi^0)\,\mathcal{B}(\pi^0\to\gamma\gamma)$ & 0.7 & 0.7 \\
\midrule
Total multiplicative uncertainty & 10.7 & 12.7  \\
\bottomrule
\end{tabular}
\label{table:mult_syst}
\end{table}

The yield uncertainty $\delta_N$, obtained by adding the corresponding
statistical and systematic uncertainties in quadrature, is combined
with the multiplicative uncertainty $\delta$ using the following
formula:
\begin{equation}
(N\pm\delta_N)\times(1\pm\delta)=N\pm(\delta_N\oplus{}N\delta\oplus\delta_N\delta),
\end{equation}
where the symbol $\oplus$ denotes addition in quadrature. The
estimated Born-level cross sections and upper limits at the 90\%
confidence level (CL) set using the likelihood-ratio ordering
method~\cite{Feldman:1997qc} are presented in
Table~\ref{table:yield_XS_UL_data}.

\section{Conclusions}

We report a search for the $e^+e^-\to\eta_{b}(1S)\omega$ and
$e^+e^-\to\chi_{b}(1S)\omega$ processes at $\sqrt{s}=10.745\,\gev$. No
significant signals are observed, and we set the following 90\% CL
upper limits on Born-level cross sections:
\begin{equation}
\begin{split}
\sigma_\mathrm{B}(e^+e^- \to \eta_{b}(1S)\omega) & < 2.5\,\mathrm{pb}, \\
\sigma_\mathrm{B}(e^+e^- \to \chi_{b0}(1P)\omega) & < 8.7\,\mathrm{pb}. \\
\end{split}
\end{equation}
The upper limit on the $\ee\to\chbz\om$ cross section is comparable to
the upper limit obtained using full reconstruction of
11.3\,pb~\cite{Belle-II:2022xdi}. We combine the two results, taking
into account correlations, to obtain
\begin{equation}
\sigma_\mathrm{B}(e^+e^-\to\chi_{b0}(1P)\omega)<7.8\,\mathrm{pb}.
\end{equation}

The tetraquark model of Ref.~\cite{Wang:2019veq} predicts that the
decay rate of $\U(10753)\to\eta_{b}(1S)\omega$ is strongly enhanced
compared to the decay rates of $\U(10753)\to\U(nS)\pp$. The obtained
upper limit on $\sigma_\mathrm{B}(\et\om)$ is close to the measured
values of $\sigma_\mathrm{B}(\U(nS)\pp)$, which are in the range
$(1-3)\,\mathrm{pb}$~\cite{Belle:2019cbt}. Thus, our results do not
support the tetraquark-model prediction that the $\U(10753)\to\et\om$
decay is enhanced~\cite{Wang:2019veq}.
In the $4S-3D$ mixing model, the decay rate of
$\U(10753)\to\eta_{b}(1S)\omega$ is smaller than the decay rate of
$\U(10753)\to\U(nS)\pp$ by a factor $0.2-0.4$~\cite{Liu:2023rhh};
our upper limit is consistent with this expectation.

The upper limit on the $\chbz\om$ cross section is higher than the
measured $\chi_{b1}(1P)\omega$ and $\chi_{b2}(1P)\omega$ cross
sections of $(3.6\pm0.9)\,\textrm{pb}$ and $(2.8\pm1.3)\,\textrm{pb}$,
respectively~\cite{Belle-II:2022xdi}.
For a $4S-3D$ mixed state, the decay rate to $\chbz\om$ is expected to
be comparable to the decay rates to $\chbo\om$ and
$\chbt\om$~\cite{Li:2021jjt}; our upper limit is consistent with this
expectation.
In the charmonium sector, the decay of the $Y(4230)$ state to
$\chi_{c0}\omega$ is enhanced compared to the decays to
$\chi_{c1}\omega$ and $\chi_{c2}\omega$~\cite{BESIII:2019gjc}. We do
not find an analogous enhancement in the decay pattern of $\U(10753)$,
which may indicate that $Y(4230)$ and $\U(10753)$ have different
structures.

\section{Acknowledgements}

We are grateful to A.I.~Milstein for valuable discussions.
This work, based on data collected using the Belle II detector, which was built and commissioned prior to March 2019, was supported by
Science Committee of the Republic of Armenia Grant No.~20TTCG-1C010;
Australian Research Council and Research Grants
No.~DP200101792, 
No.~DP210101900, 
No.~DP210102831, 
No.~DE220100462, 
No.~LE210100098, 
and
No.~LE230100085; 
Austrian Federal Ministry of Education, Science and Research,
Austrian Science Fund
No.~P~31361-N36
and
No.~J4625-N,
and
Horizon 2020 ERC Starting Grant No.~947006 ``InterLeptons'';
Natural Sciences and Engineering Research Council of Canada, Compute Canada and CANARIE;
National Key R\&D Program of China under Contract No.~2022YFA1601903,
National Natural Science Foundation of China and Research Grants
No.~11575017,
No.~11761141009,
No.~11705209,
No.~11975076,
No.~12135005,
No.~12150004,
No.~12161141008,
and
No.~12175041,
and Shandong Provincial Natural Science Foundation Project~ZR2022JQ02;
the Czech Science Foundation Grant No.~22-18469S;
European Research Council, Seventh Framework PIEF-GA-2013-622527,
Horizon 2020 ERC-Advanced Grants No.~267104 and No.~884719,
Horizon 2020 ERC-Consolidator Grant No.~819127,
Horizon 2020 Marie Sklodowska-Curie Grant Agreement No.~700525 ``NIOBE''
and
No.~101026516,
and
Horizon 2020 Marie Sklodowska-Curie RISE project JENNIFER2 Grant Agreement No.~822070 (European grants);
L'Institut National de Physique Nucl\'{e}aire et de Physique des Particules (IN2P3) du CNRS
and
L'Agence Nationale de la Recherche (ANR) under grant ANR-21-CE31-0009 (France);
BMBF, DFG, HGF, MPG, and AvH Foundation (Germany);
Department of Atomic Energy under Project Identification No.~RTI 4002,
Department of Science and Technology,
and
UPES SEED funding programs
No.~UPES/R\&D-SEED-INFRA/17052023/01 and
No.~UPES/R\&D-SOE/20062022/06 (India);
Israel Science Foundation Grant No.~2476/17,
U.S.-Israel Binational Science Foundation Grant No.~2016113, and
Israel Ministry of Science Grant No.~3-16543;
Istituto Nazionale di Fisica Nucleare and the Research Grants BELLE2;
Japan Society for the Promotion of Science, Grant-in-Aid for Scientific Research Grants
No.~16H03968,
No.~16H03993,
No.~16H06492,
No.~16K05323,
No.~17H01133,
No.~17H05405,
No.~18K03621,
No.~18H03710,
No.~18H05226,
No.~19H00682, 
No.~22H00144,
No.~22K14056,
No.~22K21347,
No.~23H05433,
No.~26220706,
and
No.~26400255,
the National Institute of Informatics, and Science Information NETwork 5 (SINET5), 
and
the Ministry of Education, Culture, Sports, Science, and Technology (MEXT) of Japan;  
National Research Foundation (NRF) of Korea Grants
No.~2016R1\-D1A1B\-02012900,
No.~2018R1\-A2B\-3003643,
No.~2018R1\-A6A1A\-06024970,
No.~2019R1\-I1A3A\-01058933,
No.~2021R1\-A6A1A\-03043957,
No.~2021R1\-F1A\-1060423,
No.~2021R1\-F1A\-1064008,
No.~2022R1\-A2C\-1003993,
and
No.~RS-2022-00197659,
Radiation Science Research Institute,
Foreign Large-Size Research Facility Application Supporting project,
the Global Science Experimental Data Hub Center of the Korea Institute of Science and Technology Information
and
KREONET/GLORIAD;
Universiti Malaya RU grant, Akademi Sains Malaysia, and Ministry of Education Malaysia;
Frontiers of Science Program Contracts
No.~FOINS-296,
No.~CB-221329,
No.~CB-236394,
No.~CB-254409,
and
No.~CB-180023, and SEP-CINVESTAV Research Grant No.~237 (Mexico);
the Polish Ministry of Science and Higher Education and the National Science Center;
the Ministry of Science and Higher Education of the Russian Federation,
Agreement No.~14.W03.31.0026, and
the HSE University Basic Research Program, Moscow;
University of Tabuk Research Grants
No.~S-0256-1438 and No.~S-0280-1439 (Saudi Arabia);
Slovenian Research Agency and Research Grants
No.~J1-9124
and
No.~P1-0135;
Agencia Estatal de Investigacion, Spain
Grant No.~RYC2020-029875-I
and
Generalitat Valenciana, Spain
Grant No.~CIDEGENT/2018/020;
National Science and Technology Council,
and
Ministry of Education (Taiwan);
Thailand Center of Excellence in Physics;
TUBITAK ULAKBIM (Turkey);
National Research Foundation of Ukraine, Project No.~2020.02/0257,
and
Ministry of Education and Science of Ukraine;
the U.S. National Science Foundation and Research Grants
No.~PHY-1913789 
and
No.~PHY-2111604, 
and the U.S. Department of Energy and Research Awards
No.~DE-AC06-76RLO1830, 
No.~DE-SC0007983, 
No.~DE-SC0009824, 
No.~DE-SC0009973, 
No.~DE-SC0010007, 
No.~DE-SC0010073, 
No.~DE-SC0010118, 
No.~DE-SC0010504, 
No.~DE-SC0011784, 
No.~DE-SC0012704, 
No.~DE-SC0019230, 
No.~DE-SC0021274, 
No.~DE-SC0022350, 
No.~DE-SC0023470; 
and
the Vietnam Academy of Science and Technology (VAST) under Grants No.~NVCC.05.12/22-23 and No~DL0000.02/24-25.

These acknowledgements are not to be interpreted as an endorsement of any statement made
by any of our institutes, funding agencies, governments, or their representatives.

We thank the SuperKEKB team for the excellent operation of the
accelerator and their special efforts to accomplish the center-of-mass
energy scan that made these results possible; 
the KEK cryogenics group for the efficient operation of the detector solenoid magnet;
the KEK computer group and the NII for on-site computing support and SINET6 network support;
and the raw-data centers at BNL, DESY, GridKa, IN2P3, INFN, and the University of Victoria for off-site computing support.


\appendix

\section{Angular distributions}
\label{appendix:ang_distr}

To estimate the angular distributions of the studied processes, we use
a nonrelativistic approximation. The amplitude of the
$\ee\to\et\om(\to\pi^+\pi^-\pi^0)$ process can be written as 
\begin{equation}
    T=[\textbf{e}_1\times\textbf{p}_{\omega}]\cdot\boldsymbol{\Lambda},
\end{equation}
where $\textbf{e}_1$ is the polarization vector of the virtual photon
produced in the $e^+e^-$ annihilation,
$\textbf{p}_{\omega}=\textbf{p}_1+\textbf{p}_2+\textbf{p}_3$ is the
total momentum of the three pions measured in the c.m. frame, and
$\boldsymbol{\Lambda}$ is defined as 
\begin{equation}
    \boldsymbol{\Lambda} = E_1\,[\textbf{p}_2 \times \textbf{p}_3] -
    E_2\,[\textbf{p}_1 \times \textbf{p}_3] + E_3\,[\textbf{p}_1
      \times \textbf{p}_2], 
\end{equation}
where $E_i$ are the energies of the pions measured in the
c.m. frame. For low-momentum $\omega$, the vector
$\boldsymbol{\Lambda}$ is perpendicular to the $\omega$ decay
plane. Summation over $\textbf{e}_1$ gives the formula for magnitude
\begin{equation}
    |T|^2 = \boldsymbol{\Lambda}^2\textbf{p}_{\omega}^2 -
    (\textbf{p}_{\omega} \cdot \boldsymbol{\Lambda})^2 -
    ([\textbf{n}_1 \times \textbf{p}_{\omega}] \cdot
    \boldsymbol{\Lambda})^2,
\end{equation}
where $\textbf{n}_1$ is the unit vector along the $e^-$ beam
direction. This formula corresponds to a $1+\cos^2\theta$ distribution
in the $\omega$ production angle $\theta$ (the angle between the beam
direction and the $\omega$ momentum in the c.m.\ frame) and a
$1+\cos^2\alpha$ distribution in the angle $\alpha$ between the normal
vector to the $\omega$ decay plane and the momentum of the beams
measured in the $\omega$ rest frame.

The amplitudes for the $\ee\to\chi_{bJ}\om(\to\pi^+\pi^-\pi^0)$
$(J=0,1,2)$ processes can be written as
\begin{equation}
    \begin{split}
        T_0 =& (\textbf{e}_1 \cdot \boldsymbol{\Lambda})A_0, \\
        T_1 =& [\textbf{e}_1 \times \boldsymbol{\Lambda}] \cdot \textbf{A}_1, \\
        T_2 =& e^i_1\Lambda^jA_2^{ij},\\
    \end{split}
\end{equation}
where $A_0$, $\textbf{A}_1$, and $A^{ij}_2$ are $\chi_{b0,1,2}$ wave
functions, respectively. After summing over the $\textbf{e}_1$
projections the magnitudes are
\begin{equation}
    \begin{split}
        |T_0|^2 =& \boldsymbol{\Lambda}^2 - (\textbf{n}_1 \cdot \boldsymbol{\Lambda})^2, \\
        {|T_1|^2} =& \boldsymbol{\Lambda}^2 + (\textbf{n}_1 \cdot \boldsymbol{\Lambda})^2, \\
        {|T_2|^2} =& 7\boldsymbol{\Lambda}^2 - (\textbf{n}_1 \cdot \boldsymbol{\Lambda})^2. \\
    \end{split}
\end{equation}

\end{document}

%% file: pub032-orcid.tex
  \author{I.~Adachi\,\orcidlink{0000-0003-2287-0173}} 
  \author{L.~Aggarwal\,\orcidlink{0000-0002-0909-7537}} 
  \author{H.~Ahmed\,\orcidlink{0000-0003-3976-7498}} 
  \author{H.~Aihara\,\orcidlink{0000-0002-1907-5964}} 
  \author{N.~Akopov\,\orcidlink{0000-0002-4425-2096}} 
  \author{A.~Aloisio\,\orcidlink{0000-0002-3883-6693}} 
  \author{N.~Anh~Ky\,\orcidlink{0000-0003-0471-197X}} 
  \author{D.~M.~Asner\,\orcidlink{0000-0002-1586-5790}} 
  \author{H.~Atmacan\,\orcidlink{0000-0003-2435-501X}} 
  \author{T.~Aushev\,\orcidlink{0000-0002-6347-7055}} 
% ...
\collaboration{The Belle II Collaboration}